\documentclass[aip,jap,preprint]{revtex4-1}

\usepackage{amsmath}
\usepackage{amssymb}
\usepackage{graphicx}
\usepackage{psfrag}
\usepackage{sistyle}
\usepackage{color}

\renewcommand{\Im}{\mathrm{Im}}

\definecolor{EditLBColor}{rgb}{0.3,0.3,1}

\definecolor{EditGCColor}{rgb}{0.62,0.13,0.94}

\definecolor{EditredColor}{rgb}{1,0,0}

\newcommand{\sSi}{\textbf{Si}}
\newcommand{\sSiA}{\sSi$_{\mathrm{A}}$}
\newcommand{\sTaA}{\textbf{Ta}$_{\mathrm{A}}$}

\begin{document}

\title{Measurements of mechanical thermal noise and energy dissipation in optical dielectric coatings}

\author{Tianjun Li}
\affiliation{Universit\'e de Lyon, Laboratoire de physique, ENS Lyon, CNRS, Lyon 69364, France}
\affiliation{Department of Physics, East China Normal University - 3663 Zhongshan North Rd., Shanghai 200062, China}
\altaffiliation{Present address: Physics Department, Shaoxing University, Shaoxing, 312000, China}
\author{Felipe A. Aguilar Sandoval}
\altaffiliation{Present address: Universidad de Santiago de Chile, Departamento de F\`\i sica, Avenida Ecuador 3493, Casilla 307, Correo 2, Santiago, Chile}
\author{Mickael Geitner}
\affiliation{Universit\'e de Lyon, Laboratoire de physique, ENS Lyon, CNRS, Lyon 69364, France}
\author{Gianpietro Cagnoli}
\email[]{g.cagnoli@lma.in2p3.fr}
\author{J\'er\^ome Degallaix}
\author{Vincent Dolique}
\author{Raffaele Flaminio}
\author{Dani\`ele Forest}
\author{Massimo Granata}
\author{Christophe Michel}
\author{Nazario Morgado} 
\author{Laurent Pinard}
\affiliation{Laboratoire des Mat\'eriaux Avanc\'es (LMA), IN2P3/CNRS,  Universit\'e de Lyon, F-69622 Villeurbanne, Lyon, France} 
\author{Ludovic Bellon}
\email[]{ludovic.bellon@ens-lyon.fr}
\affiliation{Universit\'e de Lyon, Laboratoire de physique, ENS Lyon, CNRS, Lyon 69364, France}

\date{\today}

\begin{abstract}
In recent years an increasing number of devices and experiments are shown to be limited by mechanical thermal noise. In particular sub-Hertz laser frequency stabilization and gravitational wave detectors, that are able to measure fluctuations of $\SI{e-18}{m/\sqrt{Hz}}$ or less, are being limited by thermal noise in the dielectric coatings deposited on mirrors. In this paper we present a new measurement of thermal noise in low absorption dielectric coatings deposited on micro-cantilevers and we compare it with the results obtained from the mechanical loss measurements. The coating thermal noise is measured on the widest range of frequencies with the highest signal to noise ratio ever achieved. In addition we present a novel technique to deduce the coating mechanical losses from the measurement of the mechanical quality factor which does not rely on the knowledge of the coating and substrate Young moduli.
The dielectric coatings are deposited by ion beam sputtering. The results presented here give a frequency independent loss angle of $\mathrm{(4.7 \pm 0.2)\times10^{-4}}$ with a Young's modulus of 118 GPa for annealed tantala from $\SI{10}{Hz}$ to $\SI{20}{kHz}$. For as-deposited silica, a weak frequency dependence ($\propto f^{-0.025}$) is observed in this frequency range, with a Young's modulus of 70 GPa and an internal damping of $\mathrm{(6.0 \pm 0.3)\times10^{-4}}$ at $\SI{16}{kHz}$, but this value decreases by one order of magnitude after annealing and the frequency dependence disappears.
\end{abstract}

\maketitle

\section{Introduction}
\begin{figure}[!t]
\centering
\includegraphics[width=2.8in]{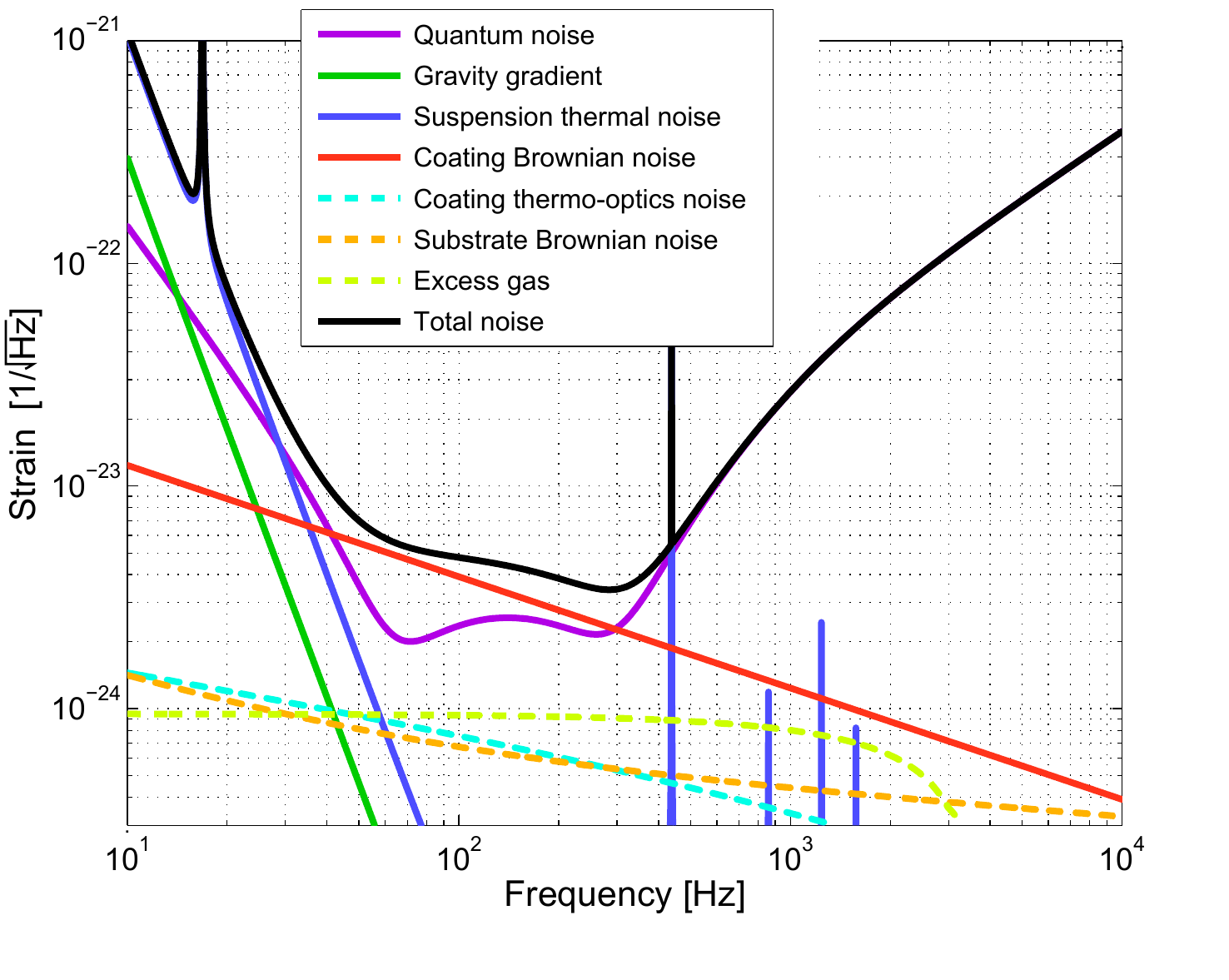}
\caption{Advanced Virgo sensitivity curve as in \cite{avirgo}. The noise level on the y axis is given in equivalent gravitational wave amplitude that has the same physical nature of strain. To convert the strain noise in displacement noise it's relevant to know that there are 4 sources of uncorrelated noise assumed to be of the same level and that the interferometer arms length is 3 km.}
\label{fig_aVirgocurve}
\end{figure}

The detection of gravitational waves is based on laser interferometry used to monitor the relative displacement of suspended masses\cite{gw}. Although the first generation of GW detectors has been concluded with an extraordinary success, the event rate at the level of sensitivity of these detectors is about one per year in the most optimistic predictions and the absence of a detected signal in the 3 years of data taking is totally compatible with the event rate uncertainty. Therefore, a second generation of detectors is being built in USA (Advanced LIGO\cite{aligo}), in Europe (Advanced Virgo\cite{avirgo}) and in Japan (KAGRA\cite{kagra}). In these advanced detectors the sensitivity limit in displacement is about $\SI{6e-21}{m/\sqrt{Hz}}$ over a wide band centred around $\SI{200}{Hz}$. The sensitivity limit of the detector Advanced Virgo, with its main noise components, is shown in Fig.\ref{fig_aVirgocurve}. Lowering the noise level by a certain factor corresponds to increase by the same factor the maximum distance at which a source can be detected. The larger this distance, the higher the event rate because the larger the number of galaxies present in the detection range. From Fig.\ref{fig_aVirgocurve} one can see that the thermal noise coming from the mirror coatings (coating Brownian noise) is the limiting noise component in the band where the detector is most sensitive. To clearly understand that, it is important to know that during the detector designing process, in general, the noise that comes from the optical readout system (Quantum noise) is shaped to match the displacement noise level. If the coating noise gets reduced the quantum noise follows, either by a simple adjustment of optical parameters or by a major redesign of the optical readout\cite{ET}, depending on the level of noise reduction to attain. 
The reflective coating on the suspend masses are made by stacks of alternate layers of two transparent materials having different refractive index. After a long process of selection two materials have been chosen for the optical coating of GW detectors: silica (SiO$_2$, low index) and titanium-doped tantala (Ti:Ta$_2$O$_5$, high index). The selection has been made on the basis of the lowest optical absorption and thermal noise level\cite{Harry, coatingadvanced}. Lowering coating thermal noise will be beneficial also for other precision measurements where high finesse optical cavities are used, such as in the development of optical clocks\cite{frequencystandard} as well as in the field of quantum opto-mechanics.

In amorphous materials thermal noise comes from unknown relaxation processes whose characteristics are fairly well explained by a model called Asymmetric Double Well Potential\cite{adwp}. In the case of bulk silica, a well known amorphous material, a number of measurements indicate that the relaxation comes from the twisting of SiO$_4$ tetrahedrons one respect to the nearest. For other materials like coated tantala, that has a higher noise level than coated silica, data are few in absolute terms, and measurements indicating the parts of the microscopic structure responsible of the relaxations are totally missing. Recent works\cite{Bassiri} have started to correlate mechanical loss of tantala to features in the reduced density function measured by electron diffraction. The results are promising.

 Direct measurement of thermal noise in multilayered coating was done on macroscopic optics for the first time in Caltech\cite{Black,Villar} and a work that aims at the thermal noise of the different components of the multistack is in preparation\cite{Villar2} by the same group. In the present work coatings are studied by observing the difference of thermal noise on micro oscillators (micro-cantilevers) before and after the coating deposition. Following Saulson~\cite{saulson} we describe the cantilever by a Simple Harmonic Oscillator (SHO) with anelastic damping: its mechanical response function $G$ linking the external force $F$ to the deflection $d$ writes in the Fourier space~\cite{f}
\begin{equation}
G(\omega) = \frac{F(\omega)}{d(\omega)} = k\left(1-\frac{\omega^{2}}{\omega_{0}^{2}}+i\phi\right)
\label{eq:G}
\end{equation}
with $\omega$ the angular frequency, $\omega_{0}$ the angular resonance frequency, $k$ the cantilever stiffness and $\phi$ its loss angle (also named mechanical loss or damping). In general $\phi$ may be frequency dependent (in this case $k$ is frequency dependent too), in the model shown here it is assumed to be constant : this is called {\it structural damping}. The quality factor of the resonance is $Q=1/\phi(\omega_{0}$). The thermal noise at room temperature $T$ is then simply obtained through the Fluctuation-Dissipation Theorem\cite{FDT} (FDT): the deflection thermal noise Power Spectrum Density (PSD) $S_{d}(\omega)$ is 
\begin{equation}
S_{d}(\omega)= - \frac{2 k_B T}{\pi \omega}\, \Im[1/G(\omega)]
\label{eq:FDT}
\end{equation}
with $k_B$ the Boltzmann's constant. Far below resonance, we immediately get the expression of the noise at low frequency :
\begin{equation}
S_{d}(\omega\ll\omega_{0})= \frac{2 k_B T}{\pi \omega}\,\frac{\phi}{k}
\label{eq_thnssimple}
\end{equation}
The signature of a structural damping is thus a $1/f$ noise at low frequency. In macroscopic bodies mechanical thermal noise is too low to be directly measured by relatively simple detectors. This is the reason why the investigation on thermal noise is almost always done indirectly through the measurement of the mechanical energy loss that is always present in the phenomenon of relaxation. When the scale of samples is reduced below 1 mm two positive effects happen: i) the elastic constant $k$ approaches $\SI{1}{N/m}$, hence thermal noise level increases up to a level ($\sim\SI{E-26}{m^{2}/Hz}$) attainable by relatively simple optical readout systems as the one used in this work; ii) the lowest resonant frequency is at several kHz, below that frequency the cantilever dynamics is equivalent to a massless spring and equation (\ref{eq_thnssimple}) holds. We will show in this article that using thermal noise of commercial AFM cantilevers, the coating internal dissipation can be measured down to a level of $\num{3E-5}$ with a good precision on a large frequency range ($\SI{10}{Hz}-$\SI{20}{kHz}).

This articles is organized as follows: Section \ref{sec:interferometer} presents the interferometer that has been used to measure directly the thermal noise on the micro cantilevers. Section \ref{sec:samples} describes the micro cantilevers that have been tested and the coatings that have been deposited. Section \ref{sec:dataanalysis} will show how the interferometric data have been analyzed to extract the relevant parameters of the oscillator. Section \ref{sec:dilution} is dedicated to the theory that have been applied to work out the relation between the coating loss angle and the one related to the entire oscillator. For the first time a technique based on the frequency shifts has been applied to give a measurement of the dilution factor and of the coating loss. Finally, in Section \ref{sec:conclusions}, the conclusions are drawn.

\section{The quadrature phase differential interferometer}
\label{sec:interferometer}

Our measurements of the micro-cantilever deflection rely on a quadrature phase differential interferometer~\cite{Schonenberger-1989,Bellon-2002,Paolino-2013}. The optical path difference is measured between the sensing and reference beams focused respectively on the free end and close to the clamping base of the cantilever (see inset of figure \ref{fig_cantilever}). The use of the light polarization allows extending the linear range of the instrument from the sub-wavelength range, typical of a Michelson interferometer, to several micrometers. In this way one can avoid the control of the reference mirror position to find the optimal working point (middle fringe position). Moreover, by design the interferometer is nearly common path, resulting in a low sensitivity to external vibration and a low drift, allowing precise measurements of mechanical thermal noise. \\
The total light intensity on the cantilever is less than $\SI{500}{\mu W}$ at $\SI{633}{nm}$, resulting in a negligible heating with respect to room temperature (less than $\SI{5}{K}$). The background noise of the instrument is measured by reflection of light on a macroscopic rigid mirror, tuning light intensity on the photodiodes to the same values measured during the thermal noise measurement on the cantilevers. As seen in figure \ref{fig_PSDsilica}, this background noise $S_{bg}$ is as low as $\SI{E-27}{m^{2}/Hz}$ at high frequency, it mainly results from the shot noise of the photodiodes. At lower frequency, $1/f$ noise from the electronics is present, with a corner frequency around $\SI{100}{Hz}$. \\
More details about the instrument can be found in reference~\cite{Paolino-2013}. This set-up has notably been used previously to characterize mechanical thermal noise and viscoelastic behavior of metallic coatings on micro-cantilevers~\cite{Paolino-2009,Li-2012}.

\section{The sample preparation and parameters}
\label{sec:samples}

\begin{figure}[!t]
\centering
\includegraphics[width=2.9in]{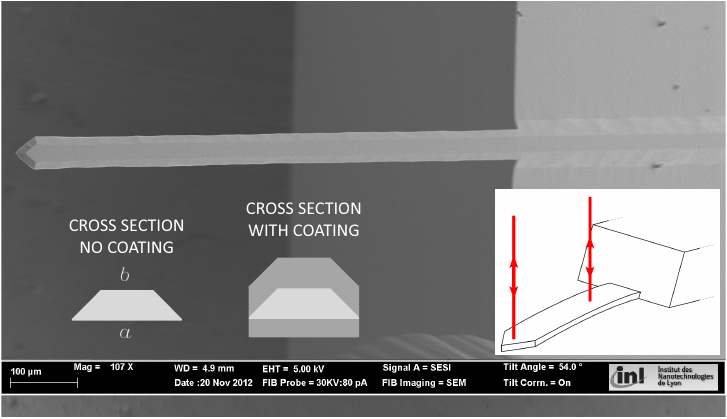}
\caption{Electron microscope image of a cantilever without coating and later used in the measurements. The length of the cantilever protruding from the silicon block is about $\SI{505}{\micro m}$. The photo should show the irregularity of the width and of the slanted faces due to the etching process. In the inset: position of the laser beams relative to the cantilever.}
\label{fig_cantilever}
\end{figure}

The samples are made of tipless AFM cantilevers\cite{cantilevercompany} with thin films of silica $\mathrm{SiO_2}$ or tantala $\mathrm{Ta_2O_5}$ deposited through Ion Beam Sputtering (IBS). A cantilever used in this experiment is shown in Fig.\ref{fig_cantilever}.

\begin{figure}[!t]
\centering
\includegraphics[width=3.5in]{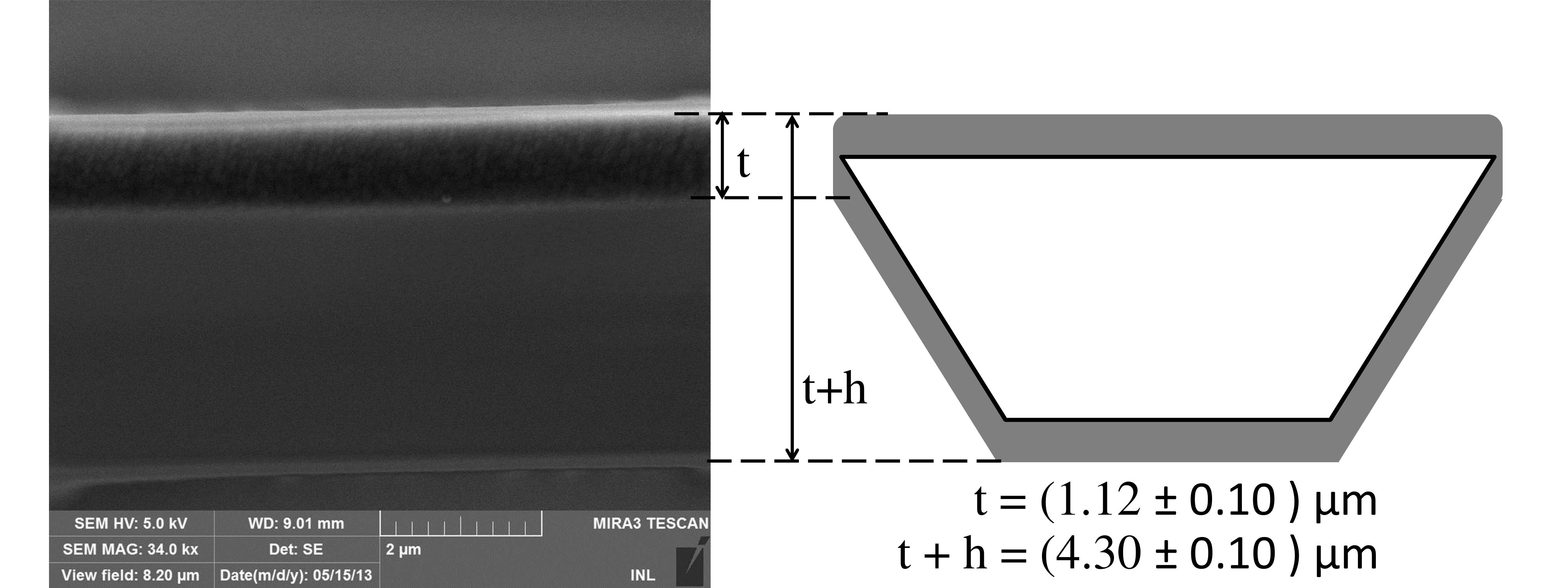}
\caption{Electron microscope image of the sample \sSiA: side view at the middle of the cantilever. Measurement of the total coating thickness $t$ (sum of the coatings deposited on the narrow and wide sides) and total cantilever thickness $t + h$ are shown.  $t$ and $h$ are respectively 12\% and 2\% higher than the values that come from the parameters reported in Table \ref{table_samples}. At right: sketch of the coated-cantilever cross section that helps to understand the electron microscope image.}
\label{fig_sideview}
\end{figure}

The cross section has the shape of a isosceles trapezium with dimensions $a=(31.4\pm0.8)\,\SI{}{\micro m}$ (the long side), $b=(16.3\pm0.9)\,\SI{}{\micro m}$ (the short side) both measured by the electron microscope. The thickness is $h_{\mathrm{Ta}}=(3.07 \pm 0.12)\,\SI{}{\micro m}$ and $h_{\mathrm{Si}}=(3.13 \pm 0.12)\,\SI{}{\micro m}$ for the cantilever coated with tantala and silica respectively, as measured indirectly through the resonant frequencies (from a linear fit between the frequencies of the firsts 3 modes and the mode number squared). Before the coating deposition a side view of the cantilever with the electron microscope was not possible and the slanted lateral surfaces made the direct measurement of thickness very inaccurate. The total length of the cantilever is $L=(505\pm5)\,\SI{}{\micro m}$. \\
Although the uniformity of the thickness of the thin film is below $\SI{1}{\%}$ over $\SI{10}{cm}$ of substrate, during the deposition the cantilever bent considerably under the coating stress and this might have an effect on the thickness uniformity. In a second imaging session, using a different microscope, a measurements of coating thickness was done on the two sides of the cantilever at different points along its length. A side view of the cantilever is shown in Fig \ref{fig_sideview}. The measurements show that the coating thickness is uniform within the reading uncertainty of 10\%. In order to have a small residual curvature the thickness of the deposited coating has to be different for the two sides of the cantilever because the cross section has the form of a trapeze. With the thicknesses listed in table \ref{table_samples} the inclination of the tip was less than $\ang{0.5}$ with respect to the cantilever base. 
The silica coating could be measured right after deposition (sample \sSi), and after annealing $\SI{10}{h}$ at $\SI{500}{\degC}$ (sample \sSiA). The cantilever used for the tantala film had a thermal oxide of $20-\SI{30}{nm}$ grown before deposition. The parameters of the heat treatment are:  heating rate of $\SI{100}{\degC/h}$ up to $\SI{800}{\degC}$ and then after 1 minute cooled to room temperature at the same rate. This oxide layer is necessary for the adhesion of tantala. After deposition the tantala sample \sTaA\ was annealed for $\SI{10}{h}$ at $\SI{500}{\degC}$.

\begin{table}[!h]
\caption{Coating parameters for the two measured samples.}
\label{table_samples}
\centering
\begin{tabular}{|l|c|c|c|c|c|}
\hline
Sample		& Coating		& Thickness			& Thickness			& $\rho$			& E 		\\
			& material		& on $a$ ($\SI{}{nm}$)	& on $b$ ($\SI{}{nm}$) 	& ($\SI{}{g/cm^3}$)	& ($\SI{}{GPa}$)	\\
\hline
\sSi 			& SiO$_2$		& $424 \pm 4$ 		& $541 \pm 5$		& $2.52 \pm 0.02$		& 70		\\
\hline
\sSiA\			& SiO$_2$		& $432 \pm 4$ 	 	& $552 \pm 5$		& $2.47 \pm 0.02$		& 70		\\
\hline
\sTaA 			& Ta$_2$O$_5$	& $378 \pm 4$		& $482 \pm 5$		& $7.39 \pm 0.02$		& 140 		\\
\hline
\end{tabular}
\end{table}
The density of tantala and silica were measured by mass measurement on silicon wafers. The annealing causes a decrease of coating density that correspond to an increase of coating thickness. The thicknesses reported in Table \ref{table_samples} are corrected by this effect from the nominal values coming from the deposition process. The Young's moduli reported in this table are compatible with values measured by nanoindentation \cite{Abernathy}. Two cantilevers have been coated in total.

\section{Data processing of the interferometer signals and results.}
\label{sec:dataanalysis}

The cantilever is placed at room temperature in a vacuum chamber to reduce the viscous damping by the surrounding atmosphere (static vacuum around $\SI{3e-2}{mbar}$). Its thermal noise driven deflection is measured directly by the quadrature phase interferometer. The data acquisition system samples the signals at a rate of $\SI{250}{kHz}$ with 24 bit resolution. The length of each data stream is $\SI{4}{s}$. After an off-line processing of the sampled photodiodes signals a PSD calibrated in $\,\SI{}{m^2/Hz}$ is produced. After a data cleaning process that removes all the harmonics of the power line and electronic devices, the single data stream is stored and made available for the further data analysis. Data analysis was done in three steps as described in the following sections. The cantilever loss angle is estimated at first from the shape of the resonant peak and then on a much wider frequency range using the Fluctuation-Dissipation theorem and the Kramers-Kronig relation. 

\subsection{Evaluation of the SHO parameters}

On a narrow $140\,$Hz frequency interval around the first resonance, the measured PSD is fitted with the following function~\cite{saulson}, derived from equation~(\ref{eq:FDT}):
\begin{equation}
S_{fit}(\omega)=\frac{2 k_B T}{\pi k \omega}\frac{\phi_{0}}{\left[1-\left(\omega/\omega_0\right)^2\right]^2+\phi_{0}^2} + S_{fitbg}
\label{eq_fittingfunction}
\end{equation}
where $S_{fitbg}$, the background noise level that is constant for frequencies greater than $\SI{1}{kHz}$, is measured directly from the spectra, while the angular resonant frequency of the mode $\omega_0=2\pi f_{0}$, the loss angle $\phi_{0}$ and the elastic constant $k$ are determined by the fitting process. In particular, $f_0$ is related to the peak frequency position, $\phi_{0}$ to the peak width and $1/k$ to the curve integral~\cite{k}. The temperature is assumed to be the room temperature $T=\SI{298}{K}$ since, given the laser power used during this measurement the cantilever heating is negligible. In the fitting process of a linear PSD spectrum the relative weight of the few points around the resonance is enormous in comparison to the weight of the many points far from the resonance so, in order to better balance the weight of all the points, the fitting process is done on the logarithm of the PSD. 

\begin{figure}
\centering
\includegraphics{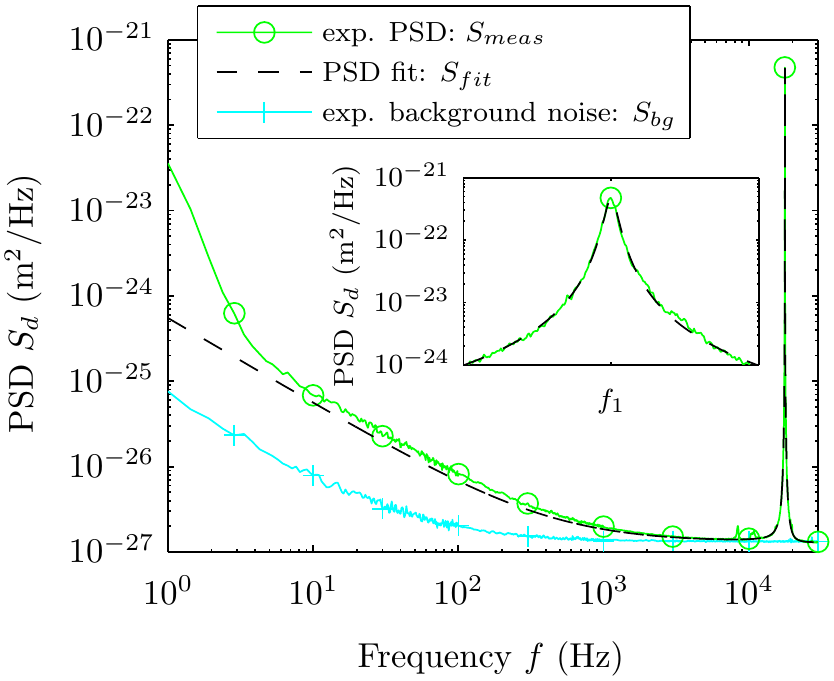}
\caption{Averaged PSD of measured deflection for 140 data sets for a silica coated cantilever measured in vacuum (sample \sSi). The curve is compared to the fitting function (\ref{eq_fittingfunction}) and the to background noise, measured on a rigid mirror. Inset: zoom on a $100\,$Hz frequency interval around resonance.}
\label{fig_PSDsilica}
\end{figure}

For each thermal noise measurement done in this work, at least 120 spectra were taken. In order to deal with any potential non stationary process two kinds of average were done: the average of each parameter that results from the fitting of each single spectrum and the fitting done on the average spectra. In all the measurements done in this work the two averages gave the same results within the experimental uncertainties. The averaged PSDs $S_{meas}$ and the fitting curves $S_{fit}$ are given in Fig.\ref{fig_PSDsilica} for sample \sSi. In Table \ref{table_fittingparameters} are listed the results of the fitting process on spectra taken before and after the coating deposition for the three measured samples~\cite{k}. The effect of annealing on the silica sample is clearly evidenced by the very low internal damping measured on sample \sSiA.
\begin{table}[!h]
\caption{Fitting parameters~\cite{k}. The uncertainty corresponds to the statistical error on the fits of more than 120 spectra (only 20 for uncoated samples).}
\label{table_fittingparameters}
\centering
\begin{tabular}{|l|l|l|}
\hline
Sample & $f_0$ ($\,\SI{}{Hz}$) & $\phi_{0} / 10^{-4}$ \\
\hline
Uncoated \sSi\ \& \sSiA & $16501 \pm 2$	& $0.3\pm0.3$ \\
\hline
Sample \sSi & $17732.2 \pm 0.1$ & $2.59\pm0.04$	\\
\hline
Sample \sSiA & $17835.2 \pm 0.1$	& $0.43\pm0.02$	\\
\hline
Uncoated \sTaA & $16110 \pm 1$ & $0.3\pm0.3$ \\
\hline
Sample \sTaA & $14771.6 \pm 0.1$ & $2.28\pm0.03$	\\
\hline
\end{tabular}
\end{table}

\subsection{Extraction of the cantilever thermal noise PSD}

\begin{figure}
\centering
\includegraphics{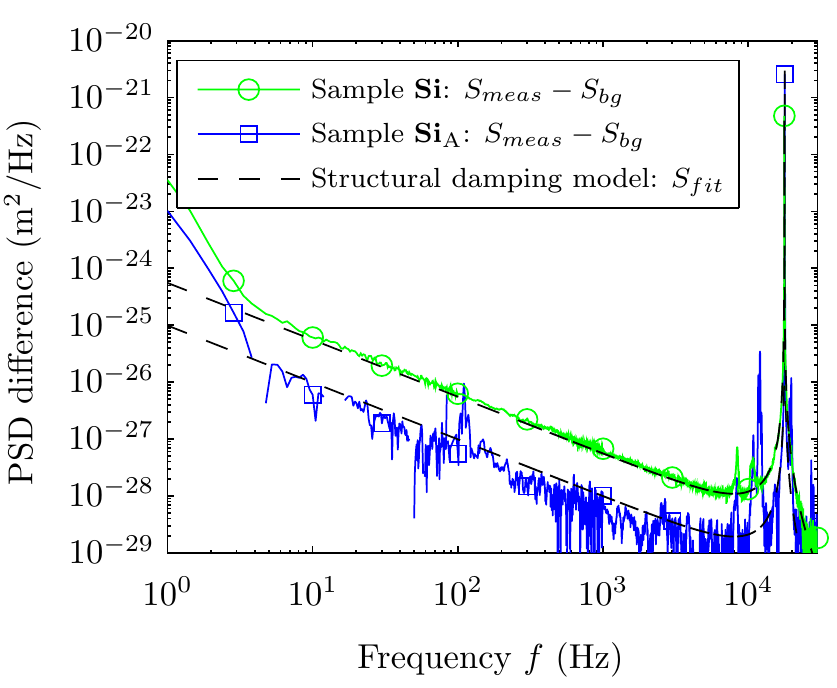}
\caption{PSD of samples \sSi\ and \sSiA\ coated with silica obtained by subtraction of the averaged PSD of measured signals $S_{meas}$ and averaged PSD of background noise $S_{bg}$ (PSDs of figure~\ref{fig_PSDsilica} for sample \sSi). The fit of the resonance with equation (\ref{eq_fittingfunction}) is still valid at low frequencies, demonstrating the very weak frequency dependence of $\phi$. The thermal noise of the annealed sample is an order of magnitude lower than that of the as-coated sample, almost reaching the limit of our detection system. Peaks around $\SI{10}{kHz}$ and high noise level below $\SI{5}{Hz}$ are residual environmental noise.}
\label{fig_PSDsilicasub}
\end{figure}

\begin{figure}[!h]
\centering
\includegraphics{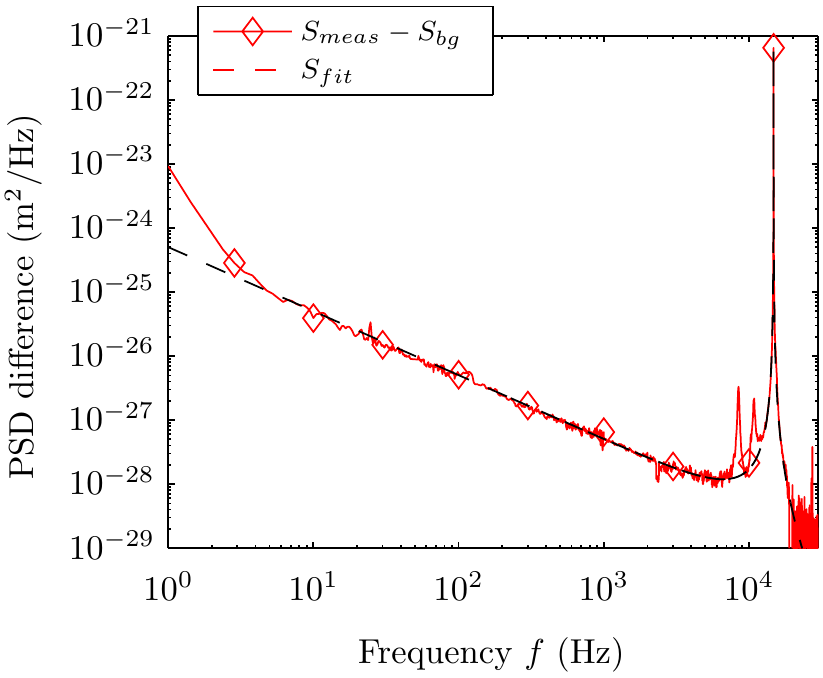}
\caption{PSD of sample \sTaA\ coated with tantala obtained by subtraction of the averaged PSD of measured signals $S_{meas}$ and averaged PSD of background noise $S_{bg}$. The fit of the resonance with equation (\ref{eq_fittingfunction}) is still valid at low frequencies, demonstrating the frequency independence of $\phi$. Peaks around $\SI{10}{kHz}$ and high noise level below $\SI{5}{Hz}$ are residual environmental noise.}
\label{fig_PSDtantalasub}
\end{figure}

In the thermal noise spectrum of Fig.\ref{fig_PSDsilica} a clear $1/f$ noise can be seen at low frequency. Although a contribution from the instrument background noise is present, the measured signal is much larger than this detection limit. The mechanical thermal noise contribution of the sample can be extracted by the difference between the averaged PSD of the measured signal $S_{meas}$ and the averaged PSD of the background noise $S_{bg}$ (measured on a rigid mirror). The result of this subtraction is given in Fig.\ref{fig_PSDsilicasub} and Fig.\ref{fig_PSDtantalasub} for the silica and tantala coated samples respectively. At frequencies lower than $\SI{5}{kHz}$ the noise is proportional to $1/f$ as predicted by the structural noise model with constant loss eq.(\ref{eq_thnssimple}). The fit reported in those figures is simply the one obtained previously from eq.(\ref{eq_fittingfunction}) (imposing $S_{fitbg}=0$ in this case): its parameters are adjusted in a $140\,$Hz frequency range around the resonance. We see however that the agreement at any frequency is good, demonstrating the weak dependence on $f$ of the structural damping $\phi$.

\subsection{Extraction of the cantilever internal dissipation $\phi$}

\begin{figure}[!h]
\centering
\includegraphics{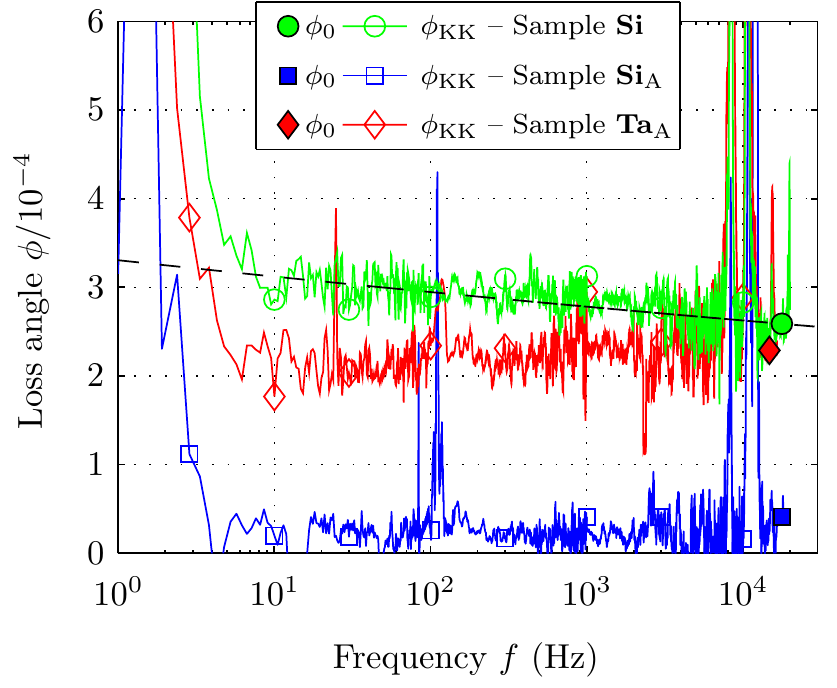}
\caption{Internal damping $\phi$ as a function of frequency for the three measured samples. The result of the reconstruction process through Kramers-Kronig's relations $\phi_{\mathrm{KK}}$ is in very good agreement with the structural damping model ($\phi_{0}$ measured with the fitting procedure of the thermal noise at resonance through equation~(\ref{eq_fittingfunction})). The dissipation is much weaker after annealing for the silica coating, reaching the limits of our detection system. Peaks around $\SI{10}{kHz}$ and high noise level below $\SI{5}{Hz}$ are due to the residual environmental noise.}
\label{fig_phi}
\end{figure}

Since we measure the PSD of thermal fluctuations on a wide frequency range, we can use the FDT and Kramers-Kronig's relations to rebuilt the full mechanical response $G(\omega)$ of the cantilever~\cite{Li-2012,Paolino-2009}. Indeed, through the FDT we actually measure $\Im[1/G(\omega)]$ (see equation~(\ref{eq:FDT})). We then apply Kramers-Kronig's integral relations for the response function $1/G(\omega)$ to infer its real part from the knowledge of its imaginary part, eventually getting a full measurement of $G(\omega)$. In figure~\ref{fig_phi} we plot the result of this reconstruction process with the imaginary part of the response function scaled by the cantilever stiffness: using equation~(\ref{eq:G}), this should directly lead to the internal damping of the cantilever $\phi=\Im(G)/k$. We notice that the frequency dependance is very weak for the 3 samples: $\phi(\omega)$ is flat between $\SI{10}{Hz}$ and $\SI{20}{kHz}$ for the annealed samples \sSiA\ and \sTaA, and presents a very slight decrease for sample \sSi. This weak frequency dependence can be fitted for example with a power law $\omega^{\alpha}$, with $\alpha=-0.025$.

\section{From the cantilever dissipation to the coating loss angle}
\label{sec:dilution}

In this section the relation between the loss angle of the resonator $\phi$ and the loss angle of the coating $\phi_{\mathrm{c}}$ will be worked out. To do that one can use the definition of quality factor knowing that at the resonant frequencies $\phi=1/Q$. The extention of this relation to all frequencies is assumed. Knowing that $Q = 2\pi\,E/E_{\mathrm{lost}}$, where $E_{\mathrm{lost}}$ is the energy lost in one cycle, $E$ is the total energy of the oscillator, and assuming only two components (the substrate s and the coating c), the cantilever loss angle is written:  
\begin{equation}
\phi\,=\,\frac{E_{\mathrm{lost\,c}}+E_{\mathrm{lost\,s}}}{2\pi\;E}\,=\,\frac{\phi_{\mathrm{c}}2\pi\,E_{\mathrm{c}}+\phi_{\mathrm{s}}2\pi\,E_{\mathrm{s}}}{2\pi\;E}\,=\,\phi_{\mathrm{c}}\,\mathcal{D}+\phi_{\mathrm{s}}\left(1-\mathcal{D}\right)
\label{eq:phidilution}
\end{equation}
In the previous expression the relation between energy loss and $\phi$ is applied twice, for the substrate and the coating and, finally, the {\it dilution factor} $\mathcal{D}$ has been defined as $E_{\mathrm{c}}/E$. $E_{\mathrm{c}}$ and $E_{\mathrm{s}}$ are the energy stored in the coating and in the substrate respectively, $E=E_{\mathrm{c}}+E_{\mathrm{s}}$. \\
In order to calculate the dilution factor it is assumed that the oscillator is at its maximum displacement; therefore the total energy is equivalent to the maximum potential energy that, for an unidimensional elastic beam, is:
\begin{equation}
E = V_{\mathrm{max}} = YI\int_0^L \left[\frac{\partial^2 w}{\partial z^2}\right]^2\;dz
\label{eq:pot_en}
\end{equation}
where $Y$ is the Young's modulus, $I$ the cross moment of inertia calculated with respect to the neutral line ($Y I$ is called also the beam rigidity), $w$ is the transversal displacement of the neutral line with respect to its position at rest and $z$ is the coordinate along the beam of length $L$. When equation (\ref{eq:pot_en}) applies to the coating and to the substrate, only the rigidity makes the difference between the two energies because the mode shape $w(z)$ is the same for both. Hence, the dilution factor becomes:
\begin{equation}
\mathcal{D}\,=\,\frac{V_{\mathrm{c}}}{V}\,=\,\frac{Y_{\mathrm{c}}I_{\mathrm{c}}}{Y_{\mathrm{c}}I_{\mathrm{c}}+Y_{\mathrm{s}}I_{\mathrm{s}}^{\mathrm{c}}}
\label{eq:dil_def}
\end{equation}
In the previous expression it appears $I_{\mathrm{s}}^{\mathrm{c}}$ which is the substrate cross moment of inertia when the coating is present. In facts, the uncoated-substrate cross-moment of inertia $I_{\mathrm{s}}^{\mathrm{u}}$ is different from the previous one because the position of the neutral line may change when the coating is applied. \\
Using the dimensions reported in Section \ref{sec:samples} we have calculated the beam rigidities for the various components in the two measured sampled, assuming $70$ GPa, $140$ GPa and $169$ GPa the Young's modulus of coated silica, coated tantala and silicon along the direction $<110>$, respectively. The expression of the rigidities with all the parameters is rather complicated and it is not of any particular use to have it written here. Therefore we report only the results of the dilution factor calculations: $\mathcal{D}_{\mathrm{Si}}^{calc} = 0.41$ and $\mathcal{D}_{\mathrm{Ta}}^{calc} = 0.54$. \\
On measuring the frequency shift of the resonant modes before and after the coating deposition one can have a way to measure directly the dilution factor. In facts the resonant angular frequencies $\omega_n$ are related to the potential energy $V_n$ of each mode $n$ through the equivalence $V_n = K_n$ where $K_n$ is the kinetic energy. In an explicit form the previous equation reads:
\begin{equation}
V_n = \omega_n^2\,\mu\,\int_0^L \left[w_n(z)\right]^2\;dz
\label{eq:energy}
\end{equation}
where $\mu$ is the linear mass density defined as the integration of density $\rho$ over the beam cross section. The previous equation is fulfilled independently of a multiplicative factor (the amplitude of the mode) of $w_n(z)$. The energy equation (\ref{eq:energy}) written for the uncoated and the coated beams follows:
\begin{align}
V_{s,n}^u &= (\omega_n^u)^2\,\mu_s\,\int_0^L \left[w_n(z)\right]^2\;dz \label{eq:energyuncoated} \\
V_{s,n}^c + V_{c,n}^c &= (\omega_n^c)^2\,(\mu_s + \mu_c)\,\int_0^L \left[w_n(z)\right]^2\;dz 
\label{eq:energycoated} 
\end{align}
Dividing term by term the previous equations and remembering the definition of potential energy (\ref{eq:pot_en}) and dilution factor (\ref{eq:dil_def}) a simple relation between dilution factor and frequency shifts is obtained:
\begin{equation}
\frac{I_s^u}{I_s^c}\,\left[ 1 - \mathcal{D} \right] = \left(\frac{f^u}{f^c}\right)^2 \, \frac{\mu_s}{\mu_s + \mu_c}
\label{eq:pippo}
\end{equation}
On cantilevers and wafers substrates that are bent by the action of the stress in the coating the previous relation does not apply. The reason of that is still unclear and it is being investigated by several research groups. Our samples have the coating deposited on both sides and the cantilever is almost straight (the tip angle is only $\ang{0.5}$ with respect to the cantilever base).
In the previous expression only one quantity (that needs to be calculated, being impossible to measure it) still depends on the Young's modulus of coating: $I_s^c$. However for the cantilevers and coatings used in this study, its difference with $I_s^u$ is relatively small, of the order of one part in $10^4$, so the ratio ${I_s^u}/{I_s^c}$ is from now on considered equal to 1. 
With this assumption, equation (\ref{eq:pippo}) shows the link of the dilution factor to quantities that are easily measurable and that are not related to the knowledge of the elastic constants of coating. 
Contrary to previous works\cite{Berry} where the authors gave a linearized version of the equation (\ref{eq:pippo}) valid for coatings much thinner as compared to the substrates, in our experiment the measurement of the dilution factor or of the coating elastic constant by frequency shift seems to be reliable, probably due to the small difference between coating and substrate thickness.
Taking the dimensions of samples in Section \ref{sec:samples} and the density worked out by mass measurements on silicon wafers, one can calculate the dilution factors from the measured frequency shift as reported in the table \ref{tab:dilutions}.
\begin{table}[!h]
\renewcommand{\arraystretch}{1.3}
\caption{The measured vs. calculated dilution factors.}
\label{tab:dilutions}
\centering
\begin{tabular}{|| l | c | c | c |  c | c | c ||}
\hline
Sample		& $f^u$ [Hz]		& $f^c$ [Hz]			& $\mu_s\times 10^{8}$	& $\mu_c\times 10^{8}$	& $\mathcal{D}$	& $\mathcal{D}^{\mathrm{calc}}$	\\
                                &                              &                                        & [g/cm]                             & [g/cm]			&			&						\\
\hline
 Sample \sSi		& $16501 \pm 2$	& $17732.2 \pm 0.1$	& $174 \pm 9$		& $76 \pm 2$		& $0.40 \pm 0.02$	& $0.40 \pm 0.02$			\\
 Sample \sSiA	& $16501 \pm 2$	& $17835.2 \pm 0.1$	& $174 \pm 9$		& $76 \pm 2$		& $0.41 \pm 0.02$	& $0.41 \pm 0.02$			\\
 Sample \sTaA	& $16110 \pm 1$	& $14771.6 \pm 0.1$	& $171 \pm 9$		& $200 \pm 5$		& $0.45 \pm 0.02$	& $0.54 \pm 0.02$			\\
\hline
\end{tabular}
\end{table}
The measured and calculated dilution factor for coated silica are equal within the experimental uncertainties whereas for tantala the difference is significant and consistent with other measurements of dilution factors of coated tantala on silicon wafers (not yet published)  done by the authors. The difference can be explained by an actual tantala Young's modulus of 118 GPa rather than 140 GPa as assumed in the dilution factor calculation. \\
Once the dilution factor is known, equation (\ref{eq:phidilution}) is used to work out the loss angle of the coating from the loss angles of the substrate $\phi_{\mathrm{s}}$ and of the whole oscillator $\phi$.
With a noise measurement on an uncoated cantilever we have estimated the substrate loss $\phi_{\mathrm{s}}= (0.3\pm 0.3) \times \num{E-4}$ and then the resulting coating losses for the two types of coatings are listed in Table \ref{table_phicoatings}.
\begin{table}[!h]
\renewcommand{\arraystretch}{1.3}
\caption{Loss angle of coatings deposited on Silicon micro cantilevers}
\label{table_phicoatings}
\centering
\begin{tabular}{|| l | c | c | c | c ||}
\hline
Coating			& $\phi \times 10^4$	& $\phi_{\mathrm{s}} \times 10^4$	& $\mathcal{D}$		& $\phi_{\mathrm{c}} \times 10^4$	\\
\hline
SiO$_2$ as-coated		& $2.59 \pm 0.04$		& $0.3\pm 0.3$				& $0.40 \pm 0.02$		& $6.0 \pm 0.3\;(\pm 0.5)$		\\
SiO$_2$ annealed		& $0.43 \pm 0.02$		& $0.3\pm 0.3$				& $0.41 \pm 0.02$		& $0.62 \pm 0.05\;(\pm 0.43)$		\\
Ta$_2$O$_5$ annealed	& $2.28 \pm 0.03$		& $0.3\pm 0.3$				& $0.45 \pm 0.02$		& $4.7 \pm 0.2\;(\pm 0.4)$		\\
\hline
\end{tabular}
\end{table}
The values in brackets are the systematic errors related to the estimation of the substrate losses. The maximum systematic error correspond to a substrate loss angle equal to zero.
\begin{table}[!h]
\renewcommand{\arraystretch}{1.3}
\caption{Coating loss angles measured with the resonant method on Silicon wafers}
\label{tab:phiwafers}
\centering
\begin{tabular}{|| l | l | c | c ||}
\hline
\multicolumn{2}{|| c |}{Coating}				& $\phi_{\mathrm{c}}\times10^{4}$	& $\phi_{\mathrm{c}}\times 10^{4}$		\\
\multicolumn{2}{|| c |}{}					& as-coated					& annealed at $\SI{500}{\degC}$		\\
\hline
Tantala	& with $\mathcal{D}$			& $11.4 \pm 0.2$				& $4.90 \pm 0.25$					\\
\cline{2-4}
		& with $\mathcal{D}^{\mathrm{calc}}$	& $8.2 \pm 0.7$				& $3.8 \pm 0.4$					\\
\hline
Silica		& with $\mathcal{D}^{\mathrm{calc}}$	& $3.9 \pm 0.4$				& ---							\\
\hline
\end{tabular}
\end{table}

\section{Conclusions}
\label{sec:conclusions}
Direct measurement of thermal noise on micro cantilevers coated with tantala and silica have been performed in the frequency range from $\SI{10}{Hz}$ to $\SI{20}{kHz}$. The measurement were used to characterize the noise properties of the two amorphous materials deposited by IBS. The characterizing parameter used here is the loss angle $\phi$ that allows the calculation of thermal noise in any system where the same coating is deposited. The loss angle of annealed $\SI{500}{\degC}$ tantala and silica are independent of frequency whereas as-deposited silica shows a slight dependence on frequency as $f^{-0.025}$. The measurements of coating loss done with direct measurement of thermal noise were compared with measurements on some 3" diameter silicon wafers using the resonant method. The wafers were coated using the same coater and the same coating parameters as for the micro cantilevers. The dilution factors were both calculated and measured. The results of these measurements are shown in table \ref{tab:phiwafers}.

The loss angle measurements made by thermal noise method are in a fairly good agreement with the values obtained by the resonant method. In particular, the values obtained for tantala are in agreement within the experimental uncertainties. For as-coated silica the comparison shows a significant disagreement that might be explained by a large variability on the optical and mechanical parameters that has been observed by the authors on the as-coated samples. The annealing process, beside lowering both optical and mechanical losses, contributes to stabilize their values.

A comparison can be made also with results\cite{glasgow1},\cite{glasgow2} of coating deposited on silicon substrates and measured by the resonant method obtained by other authors. Their results for the loss of tantala annealed at different temperatures are in the range $\num{2.7E-4}$ to $\num{5E-4}$. The direct measurement of thermal noise done on micro cantilever as compared to the resonant method applied on larger substrates offer a smaller statistical uncertainty on the measurement of the loss angle and the possibility to cover continuously more than 3 decades of frequency. At the moment the major uncertainty comes from the systematic error related to the poor knowledge of the uncoated substrate noise. A lower residual gas pressure and greater statistics will lower this uncertainty.

\section*{Acknowledgment}

The authors would like to thank the Engineer Armel Descamps of the Lyon Nanotechnology Intitute (INL) for the electron microscope analysis. The authors are grateful to the LABEX Lyon Institute of Origins (ANR-10-LABX-0066) of the Universit\'e de Lyon for its financial support within the program "Investissements d'Avenir" (ANR-11-IDEX-0007) of the French government operated by the National Research Agency (ANR). This work has been also partially supported by the F\'ed\'eration de Physique Amp\'ere. Finally, T.J. Li acknowledges the financial support from the Chinese Scholar Council (CSC).


\begin{thebibliography}{20}
\bibitem{f}
In this article we use the natural frequency $f$ for experimental data and angular frequency $\omega=2\pi f$ for formulas.
\end{thebibliography}
\end{document}